%% file: main.tex
\documentclass[conference]{IEEEtran}

\usepackage{cite}
\usepackage{amsmath,amssymb,amsfonts}
\usepackage{algorithmic}
\usepackage{graphicx}
\usepackage{textcomp}
\usepackage{xcolor}
\usepackage{bm}
\usepackage{multirow}
\usepackage{booktabs}
\usepackage[nolist]{acronym}

\input{abbreviation}

\begin{document}

\title{Semantic-Aware Neural Video Codec for Error-Resilient Low-Latency Transmission}

\author{\IEEEauthorblockN{
Matin Mortaheb, Homa Esfahanizadeh,  Jinfeng Du and Harish Viswanathan
}
\IEEEauthorblockA{Nokia Bell Labs, Murray Hill, NJ 07974, USA\\
\{matin.mortaheb, homa.esfahanizadeh, jinfeng.du, harish.viswanathan\}@nokia-bell-labs.com}}

\maketitle

\begin{abstract}
Emerging physical AI systems require low-latency, task-oriented video communication over unreliable channels. We propose a semantic-aware multi-level neural video coding method for robust low-latency video transmission over unreliable channels that are abstracted as multi-level packet erasure channels. Built upon the real-time DCVC-RT neural video codec, the proposed framework introduces a semantic- and feature-aware coding strategy that partitions encoded representations into packets carrying different levels of semantic and latent-feature importance and assigns these packets to different streams, each associated with a priority level when transmitted over unreliable communication channels. We also developed an error-resilient entropy model that removes inter-packet dependencies, allowing each packet to be decoded independently  under packet losses. The complete system is trained end-to-end over the abstracted multi-level packet erasure channels, enabling learning of channel-aware representations together with importance-aware packet assignment while facilitating the network for differentiated packet prioritization. Experiments show that the proposed framework significantly improves robustness over baseline DCVC-RT under packet erasures, achieving graceful degradation in less important regions while better preserving task-relevant visual content.
\end{abstract}

\section{Introduction and Preliminaries}

\begin{figure}[t]
    \centering
    \includegraphics[width=1\linewidth]{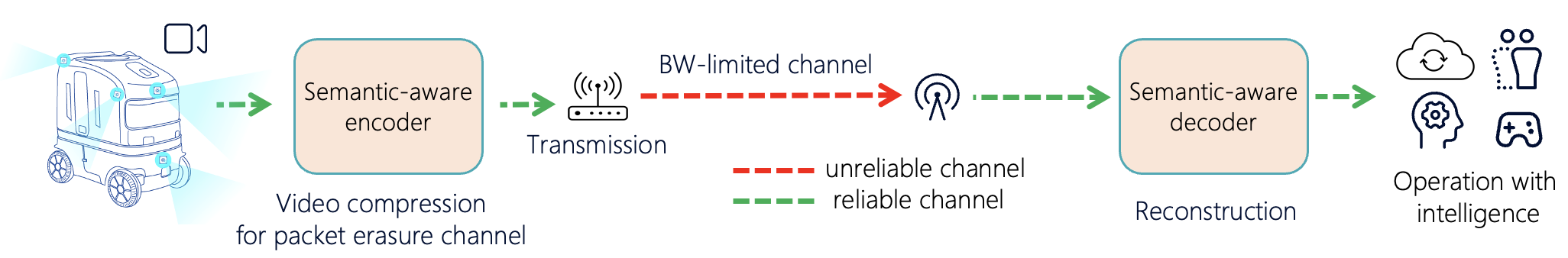}
    \caption{Task-aware error-resilient neural video codec in physical AI}
    \label{fig:introduction}
\end{figure}

Emerging physical AI systems represent a new paradigm in 6G communications, where mobile platforms such as autonomous vehicles and robotic agents interact with their physical environment and continuously transmit visual feeds or sensor data to remote servers over imperfect wireless channels \cite{Saad2020Vision6G, caballero2025physicalai}. The server must interpret this information and issue navigation or control commands in real time, making low-latency and reliable video transmission a critical requirement. However, wireless links in practical deployments are inherently unreliable because of channel fading, exhibiting packet loss and throughput fluctuations that can severely degrade transmission quality (see Fig.~\ref{fig:introduction}). In such systems, the objective is not to perfectly reconstruct every pixel of a video frame, but rather to preserve information most relevant for downstream tasks such as perception, localization, and decision-making. This motivates semantic communication frameworks, where the system prioritizes task-relevant semantic information instead of uniformly protecting all transmitted bits~\cite{DeepSem,SemcomDeniz, mortaheb2024efficient,mortaheb2025multimodal,Infoshape,TexShape}. Focusing on transmitting semantically relevant information is attractive for physical AI since it can dramatically reduce the amount of data that needs to be transmitted, thereby also increasing the likelihood of achieving low latency over capacity constrained communication links.

In such low-latency systems, end-to-end packet retransmissions or source rate adaptation based on feedback is often infeasible due to strict delay constraints. Consequently, communication systems operating in such settings must remain robust to packet erasures and channel impairments without relying on retransmission mechanisms. However, conventional video codecs such as H.264--H.266 \cite{wiegand2003h264, bross2021vvc} are designed under the assumption of near error-free transmission. When packet losses occur, entropy-coded bitstreams become highly fragile, where even small corruption may lead to catastrophic decoding failures and severe reconstruction degradation~\cite{wang2022dvst}. Furthermore, conventional codecs do not incorporate mechanisms for prioritizing semantically important regions during encoding and transmission, limiting their suitability for task-oriented communication.

Recent advances in \acp{NVC} have enabled end-to-end optimization of video compression pipelines through deep learning-based encoder, decoder, and entropy modeling components~\cite{lu2019dvc, li2021deep, dcvcrt, balle2018hyperprior}. Existing \ac{NVC} frameworks broadly fall into two categories: explicit motion compensation and implicit temporal modeling. In explicit motion compensation approaches, e.g., \ac{DVC}~\cite{lu2019dvc}, neural networks estimate motion between consecutive frames and use the resulting motion information to predict the current frame from previously reconstructed frames, transmitting only the motion parameters and residual differences. In contrast, implicit temporal modeling captures temporal dependencies directly within learned latent representations conditioned on prior frames, avoiding explicit motion estimation and often enabling more efficient real-time implementations, e.g., \ac{DCVC}-\ac{RT}~\cite{dcvcrt}. 

Despite these advances, existing \acp{NVC} remain primarily optimized for pixel-level reconstruction metrics such as \ac{PSNR} and \ac{MS-SSIM}, and are typically trained under the assumption of error-free transmission. As a result, their entropy-coded representations remain highly vulnerable to packet erasures. In particular, since entropy estimation in modern neural codecs is sequential across packets to better approximate the source distribution for improved compression efficiency, the loss of a single packet can propagate errors to subsequent packets and severely degrade quality.


Deep joint source-channel coding (DeepJSCC)~~\cite{deepwive, bourtsoulatze2019deep} addresses communication robustness by jointly training neural encoders and decoders over stochastic channel models, enabling graceful degradation of reconstruction quality under channel impairments without the cliff effect of separation-based schemes. Extensions such as layered and progressive DeepJSCC~\cite{kurka2019successive} further improve robustness through successive refinement and redundancy mechanisms. However, these frameworks are primarily designed for pixel-level reconstruction and treat all transmitted symbols uniformly, without explicitly differentiating content by semantic importance or spatial relevance to a downstream task. More recently, semantic video transmission frameworks~\cite{wang2022dvst, vista, mdvsc} have incorporated temporal priors, feature-domain context, and semantic segmentation to guide compression and transmission. While these works demonstrate the benefit of semantic awareness, they are optimized for noisy physical-layer channels modeled as AWGN or fading, and do not address packet-erasure conditions encountered at higher protocol layers, where entire packets are lost rather than corrupted by additive 
noise. In particular, none of these frameworks jointly integrates explicit region-level semantic prioritization, feature-level importance-aware packetization, and reliability-aware assignment of the packets to streams potentially experiencing different levels of packet erasures within a unified end-to-end trainable system.

A closely related work~\cite{NargisVTC, nc340417, cheng2024grace, liu2025resi} proposed block erasure-aware semantic image/video compression using a JSCC autoencoder trained over multi-level block erasure channels. In that framework, encoded packets containing more important latent features are assigned to more reliable channels. While effective for feature-level unequal error protection, the framework does not explicitly incorporate region-level semantic importance. Consequently, under packet erasures, all spatial regions may lose partial feature information, making it difficult to guarantee preservation of task-relevant regions. In addition, the framework is built upon the \ac{DVC}~\cite{lu2019dvc}, whose explicit motion compensation pipeline results in significantly slower encoding and decoding speed compared to \ac{DCVC}-\ac{RT}.

Overall, existing semantic communication and DeepJSCC frameworks primarily focus on either feature-domain robustness or semantic segmentation, often without jointly considering region-level semantics, feature importance, and transmission reliability in a unified design across heterogeneous erasure channels. In addition, existing real-time neural video codecs lack mechanisms that enable independent packet decoding under erasure conditions, limiting their robustness in practical transmission scenarios. Furthermore, most work in the literature on JSCC assume that the source encoder and channel encoder are co-located and can be jointly optimized, which is impractical for most communication systems where different entities provide the application and network.  In this paper, we propose a semantic-aware neural video coding framework built upon \ac{DCVC}-\ac{RT} for robust low-latency video transmission over unreliable channels abstracted as multi-level block erasure channels, enabling practical JSCC. Our framework jointly addresses semantic prioritization, packet-level robustness, and channel-aware transmission in a unified end-to-end trainable architecture. Specifically, our contributions are summarized as follows:

\begin{itemize}    
    \item We introduce semantic- and feature-aware packetization strategies that jointly partition encoded representations across both feature and spatial domains, enabling preferential protection of task-relevant regions.
    
    \item We propose packet assignment mechanisms according to semantic importance and latent feature significance that tag packets to different streams with potentially different reliability levels.

    \item We redesign the entropy model of DCVC-RT to enable independent entropy estimation and decoding for each packet, eliminating inter-packet dependency and preventing error propagation under packet erasures.

    \item We train the entire framework end-to-end over unreliable channels modeled as stochastic multi-level block erasure channels, enabling joint source-channel coding behavior and graceful degradation under channel impairments.
\end{itemize}

The proposed framework is compatible with emerging video transmission standards in which protocol data units (PDUs) may carry different levels of priority, enabling integration into existing transport and communication architectures.

\begin{figure*}[t]
    \centering
    \includegraphics[width=0.9\linewidth]{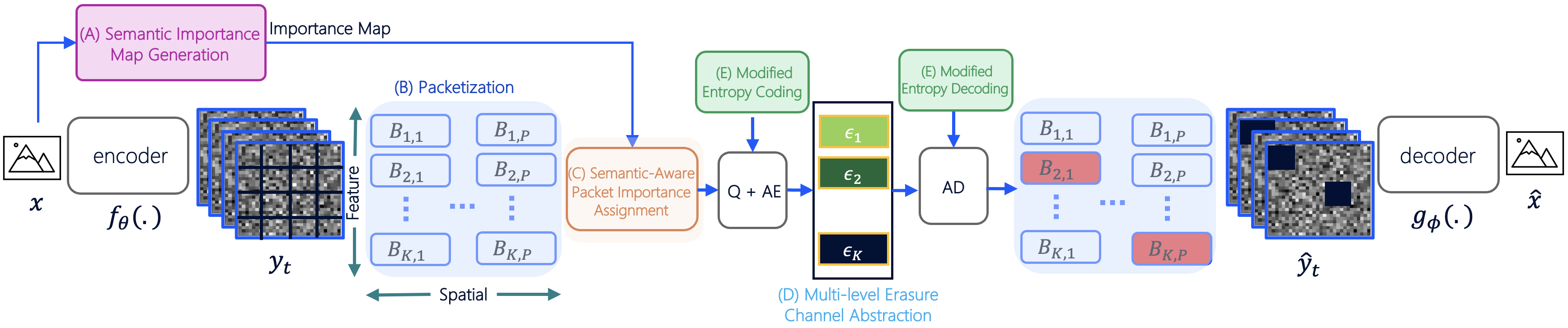}
    \caption{System model for the proposed semantic-aware neural video codec. 
The input frame $\mathbf{x}$ is encoded into a latent representation 
$\mathbf{y}_t$ by the encoder $f_\theta(\cdot)$. (A) A semantic importance 
map is generated and used to guide the packet importance assignment module. (B) the hybrid feature-region 
packetization, which partitions $\mathbf{y}_t$ into packets $B_{k,p}$ across 
both feature and spatial dimensions. (C) The importance assignment module 
allocates each packet to a protection level based on its semantic and feature 
importance. (D) Packets are transmitted through the multi-level erasure 
channel abstraction, where higher-importance packets are assigned to levels 
with lower erasure probability $\epsilon_k$; lost packets are shown in red at 
the receiver. (E) The modified entropy coding and decoding modules enable 
independent per-packet arithmetic coding and decoding, eliminating 
inter-packet dependencies and preventing error propagation. The received 
packets are reassembled into $\hat{\mathbf{y}}_t$ and passed to the decoder 
$g_\phi(\cdot)$ to reconstruct the output frame $\hat{\mathbf{x}}$.}
    \label{fig:system_model}
\end{figure*}

\subsection{Preliminaries}
The proposed system is built upon \ac{DCVC}-\ac{RT}~\cite{dcvcrt}, a low-latency neural video codec designed for real-time video compression. \ac{DCVC}-\ac{RT} encodes video frames into compact latent representations using learned encoder and decoder networks, which are subsequently entropy coded for transmission. The compression quality is controlled by a \ac{QP}, which scales the latent representation before entropy coding. A lower \ac{QP} value results in stronger quantization, producing a more compact bitstream at the cost of reconstruction fidelity, while a higher \ac{QP} value preserves higher quality at the expense of increased bitrate.
For intra-coded (I) video frames, the encoder generates a latent representation together with hyperprior side information used for entropy estimation. For predictive (P) video frames, rather than employing explicit motion compensation and residual coding, \ac{DCVC}-\ac{RT} models temporal dependencies implicitly through learned latent representations conditioned on previously decoded frames. Specifically, the decoder output of the previous frame is passed through a feature extractor module, and the resulting temporal context is concatenated with the current frame's representation at the encoder, enabling efficient exploitation of inter-frame redundancy without explicit motion estimation. At the decoder, entropy decoding and neural reconstruction networks recover the compressed frame.

Despite its compression efficiency and real-time capability, \ac{DCVC}-\ac{RT}'s original design assumes reliable, error-free communication. Its entropy model contains sequential dependencies across packets, such that the loss of a single packet can corrupt the probability estimates for all subsequent packets, leading to severe reconstruction degradation. Furthermore, the codec assigns no explicit importance to semantically relevant spatial regions, treating all content uniformly during compression and transmission. The proposed framework addresses both of these limitations by augmenting \ac{DCVC}-\ac{RT} with an error-resilient packet-independent entropy model, semantic- and feature-aware packetization, and importance-aware packet level assignment, as described in the following subsections.

\section{Method}

We propose a semantic-aware, multi-level neural video communication framework for robust, low-latency transmission of task-relevant video content over unreliable wireless channels abstracted as multi-level erasure channels. The proposed framework extends the real-time neural video codec \ac{DCVC}-\ac{RT}~\cite{dcvcrt} by introducing semantic awareness and importance-aware packet transmission, enabling joint optimization of compression, semantic prioritization, and transmission robustness within a unified end-to-end trainable architecture. As illustrated in Fig.~\ref{fig:system_model}, the system consists of five major components: (1) a neural video encoder and a semantic importance estimation module, (2) a semantic- and feature-aware packetization, (3) semantic-aware importance assignment module, (4) a multi-level erasure channel abstraction, and (5) a neural video decoder with error-resilient entropy decoding.

Given an input video sequence, the encoder compresses each frame into a compact latent representation, which is subsequently partitioned into packets according to both spatial region importance and latent feature significance (Section~\ref{sec:packetization}). In parallel, a semantic analysis module processes each frame alongside a task-relevant textual query to produce a pixel-level importance map, identifying spatial regions most critical to the communication objective (Section~\ref{sec:semantic_map}). The importance assignment module leverages both the semantic importance map and the latent feature importance to allocate packets to different protection levels, ensuring that task-relevant information receives preferential protection against erasures (Section~\ref{sec:routing}). This allocation is realized through a multi-level erasure channel abstraction that models the unreliable channel as a set of levels with varying erasure probabilities (Section~\ref{sec:channel_interface}). At the receiver, successfully received packets are entropy decoded independently and used to reconstruct the video frame, with missing packets replaced by zeros (Section~\ref{sec:decoding}). Unlike conventional neural video communication systems that treat all packets uniformly, the proposed framework explicitly aligns packet importance with protection level through end-to-end training over stochastic erasures (Section~\ref{sec:training}), enabling graceful degradation in less important regions while preserving task-relevant visual content under adverse channel conditions. Fig.~\ref{fig:system_model} shows the full system pipeline with each component labeled according to its corresponding subsection.

\subsection{Semantic Importance Map Generation}
\label{sec:semantic_map}

To enable task-oriented communication, the proposed framework employs a semantic analysis module that generates a spatial importance map for each input frame, identifying regions most relevant to a user-specified communication task and assigning higher transmission priority to those regions during importance assignment. Given an input frame $\mathbf{x}$ and a task description or textual query $q$, the semantic module produces a pixel-level importance map $\mathbf{M} \in [0,1]^{H \times W}$, where larger values indicate greater relevance to the communication objective. The module is not restricted to a specific architecture and may be implemented using an open-vocabulary segmentation model, a task-specific detector, or any vision-language model capable of estimating region relevance. In our implementation, we employ an open-vocabulary video segmentation model, ReferEverything~\cite{bagchi2025refereverything}, which accepts both a video frame and a textual query and produces a continuous relevance map corresponding to the queried content.

Rather than using the final binary segmentation mask, we utilize the continuous confidence values produced before thresholding, which provide the granularity necessary for unequal error protection. These confidence values are quantized into $N_L$ discrete semantic importance levels $S \in \{1, 2, \ldots, N_L\}$, where larger values correspond to regions more important for the communication task. In our experiments, we set $N_L = 4$. Since importance assignment is performed at the patch level, the pixel-level importance map is converted into a patch-level representation. Each frame is divided into a $4 \times 4$ grid of $N_p = 16$ spatial patches, and the importance level of patch $j$ is computed as $S_j = \mathcal{Q}\!\left( \frac{1}{|\Omega_j|} \sum_{\mathbf{x} \in \Omega_j} M(\mathbf{x}) \right)$, where $\Omega_j$ denotes the set of pixels belonging to patch $j$ and $\mathcal{Q}(\cdot)$ denotes the quantization operator. The resulting patch importance levels are used by the importance assignment module described in Section~\ref{sec:routing}.

To improve generalization and reduce training complexity, semantic maps are not generated by the segmentation model during training. Instead, randomly generated and spatially smoothed importance maps are used, exposing the encoder, decoder, and importance assignment module to a diverse set of spatial importance distributions. This prevents overfitting to a particular semantic model and allows the system to generalize effectively to unseen tasks during inference.

\subsection{Semantic-Aware Packetization}
\label{sec:packetization}

In the original \ac{DCVC}-\ac{RT} framework, the encoded latent representation is partitioned into a small number of packets, e.g., four packets for I-frames and two packets for P-frames. Each packet contains information corresponding to both a subset of latent channels and a subset of spatial locations with no overlap. Consequently, the loss of a packet simultaneously removes information associated with specific features and specific regions of the frame, leading to severe degradation in reconstruction quality.

To improve robustness against packet erasures, we redesign the packetization process and introduce a hybrid semantic-aware packetization strategy that jointly exploits latent feature importance and spatial region importance. Let the encoded latent representation of a frame be denoted by $\mathbf{Y} \in \mathbb{R}^{H \times W \times C}$, where $H$, $W$, and $C$ denote the latent height, latent width, and number of latent channels, respectively.

A natural baseline approach~\cite{nc340417} partitions the latent representation only along the feature dimension, dividing it into $N_s$ channel slices $\mathbf{Y} = \left[\mathbf{Y}^{(1)}, \mathbf{Y}^{(2)}, \ldots, \mathbf{Y}^{(N_s)}\right], \quad \mathbf{Y}^{(s)} \in \mathbb{R}^{H \times W \times \frac{C}{N_s}}, s \in \{1, 2, \ldots, N_s\}$ where each slice spans the entire spatial extent of the frame. Through end-to-end training, the encoder learns to concentrate more important latent features into slices assigned to higher protection levels, so that packet losses primarily affect lower-priority features while globally preserving the most critical information. However, since every packet spans the entire frame spatially, this strategy provides no mechanism to explicitly protect task-relevant spatial regions, limiting its effectiveness for task-oriented communication.

To address this limitation, we propose a hybrid packetization strategy that partitions the latent representation across both feature and spatial dimensions. Each frame is divided into a $4 \times 4$ grid of $N_p = 16$ spatial patches, and the latent representation is simultaneously divided into $N_s = 8$ feature slices. Each packet is associated with a unique pair of one spatial patch and one feature slice, carrying information corresponding to a specific region of the frame and a specific subset of latent features. This results in
$$N_{\mathrm{pkt}} = N_p \times N_s = 16 \times 8 = 128 \text{ packets per frame},$$
for both I-frames and P-frames.

Each packet is assigned two importance indicators: a semantic importance level $S_j$ determined by the corresponding spatial patch (Section~\ref{sec:semantic_map}), and a feature importance level determined by the corresponding latent slice. These indicators are used by the importance assignment module to allocate packets to different protection levels, ensuring that packets containing important features from semantically relevant regions receive preferential protection, while packets from less important regions and feature slices are more likely to be affected by erasures. This joint feature-region design provides finer-grained control over packet prioritization and enables preservation of task-relevant regions even under severe erasure conditions.

\subsection{Semantic-Aware Packet Importance Assignment}
\label{sec:routing}

After packetization, each packet must be assigned to one of the available importance levels. The objective of the importance assignment module is to align packet importance with its assigned protection level such that packets containing task-relevant information are more likely to survive channel erasures (see Section~\ref{sec:channel_interface} for the erasure channel abstraction).

Each packet $p_i$ is associated with two importance measures: (1) Semantic importance $S_i \in \{1, 2, 3, 4\}$, determined by the patch-level semantic importance map (Section~\ref{sec:semantic_map}). (2) Feature importance $F_i$, determined by the corresponding latent feature slice index (Section~\ref{sec:packetization}).

\begin{figure}[t]
    \centering
    \includegraphics[width=1\linewidth]{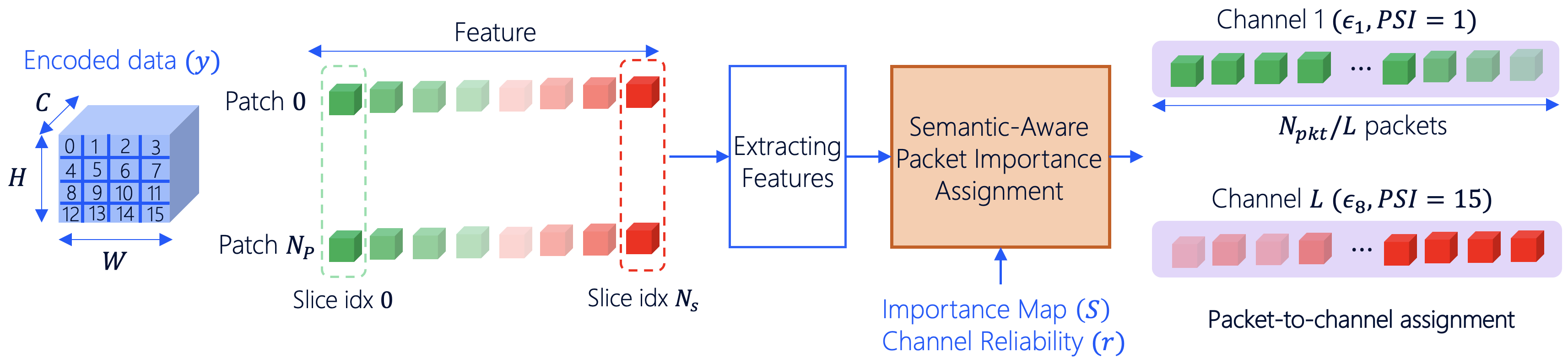}
    \caption{System model for the learned importance assignment module. The encoded data is partitioned across spatial and feature domains into $N_{pkt} = 128$ packets, where color indicates importance: green denotes high-importance packets and red denotes low-importance packets. Each packet is assigned both a slice importance score and a semantic importance score. Based on these scores, the importance assignment module allocates packets to appropriate protection levels, ensuring high-importance packets (green) are assigned to more reliable levels with lower erasure probability $\epsilon_k$, while low-importance packets (red) are assigned to less reliable levels.}
    \label{fig:packet_routing}
\end{figure}

The importance assignment problem is formulated as allocating packets to importance levels such that highly important packets are assigned to levels with greater erasure protection while maintaining balanced utilization across levels. To address this, we propose a learned importance assignment module that optimizes packet-to-level assignments jointly with the encoder and decoder through end-to-end training (see Fig.~\ref{fig:packet_routing}). The main challenge is that importance assignment involves discrete allocations, which are non-differentiable, while balanced utilization (number of packets assigned to each importance level) must be preserved. To address these constraints, we formulate importance assignment as a differentiable transport problem.

\begin{figure}[t]
    \centering
    \includegraphics[width=1\linewidth]{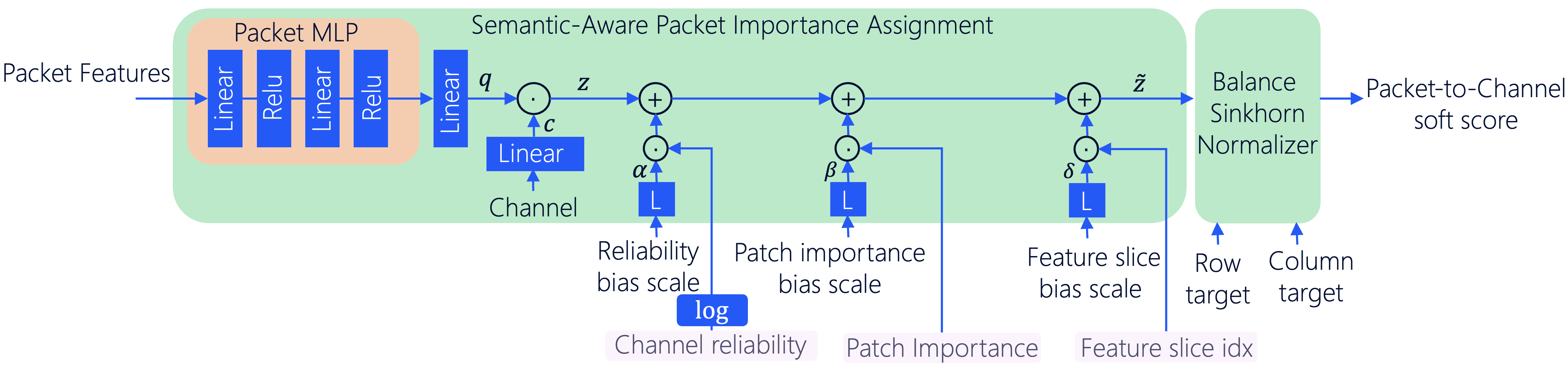}
    \caption{System model for the learned importance assignment module. The input consists of abstracted packet features, and the output is a packet-to-level soft score indicating the probability of assigning each packet to one of the available importance levels. The balanced Sinkhorn normalizer ensures that the row and column utilization targets are satisfied.}
    \label{fig:learned_routing_system}
\end{figure}

\begin{figure}[t]
    \centering
    \includegraphics[width=1\linewidth]{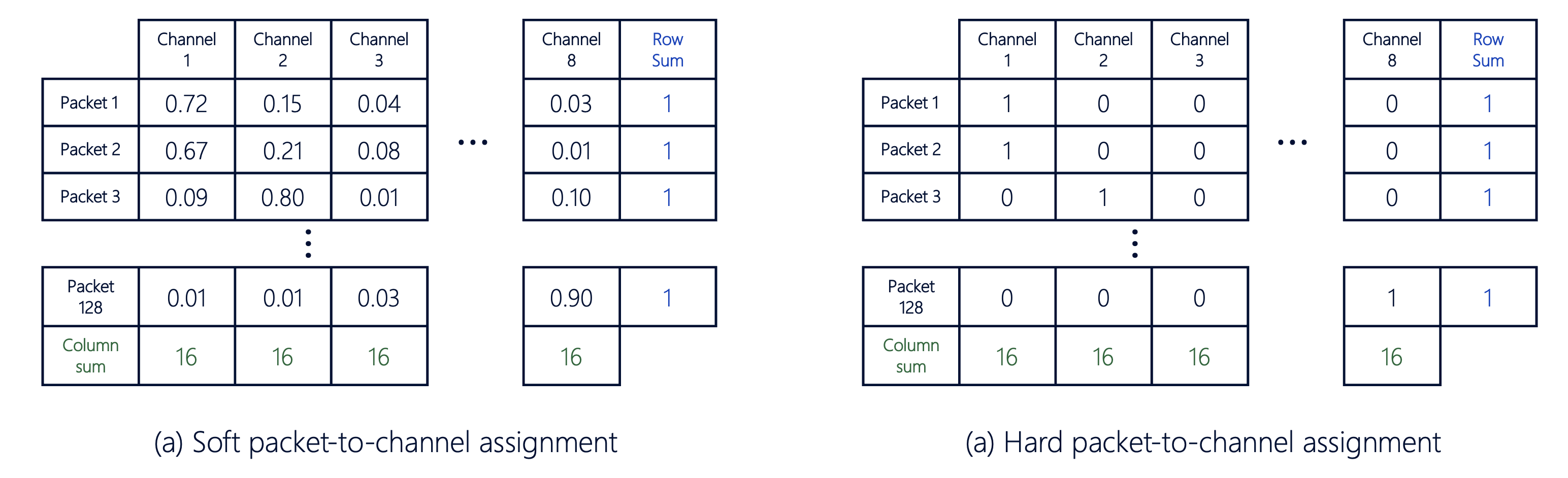}
    \caption{Packet-to-level assignment matrix produced by the learned importance assignment module: (a) soft assignment and (b) hard assignment.}
    \label{fig:packet_channel_assignment}
\end{figure}

For each packet, a set of statistical descriptors is extracted from the encoded representation, including mean value, standard deviation, and packet energy. These descriptors are passed through a lightweight neural network to produce packet embeddings $\mathbf{q}_i$. Simultaneously, each importance level is associated with a trainable embedding $\mathbf{c}_k$. Assignment logits are computed as $z_{ik} = \mathbf{q}_i^\top \mathbf{c}_k$, and subsequently modified to explicitly incorporate protection level reliability $r_k$, packet semantic importance $S_i$, and feature-slice index $F_i$:
\begin{equation}
\tilde{z}_{ik} = z_{ik} + \alpha \log(r_k) + \beta S_i + \delta F_i,
\end{equation}
where $r_k = 1 - \epsilon_k$ denotes the reliability of importance level $k$, and $\alpha$, $\beta$, and $\delta$ are learnable scalar parameters. The modified logits are processed using a balanced Sinkhorn normalization layer, producing a soft transport matrix $\mathbf{T} \in \mathbb{R}^{N_{\mathrm{pkt}} \times 8}$, whose element $T_{ik}$ denotes the probability of assigning packet $i$ to importance level $k$. The Sinkhorn normalization enforces $\sum_{k=1}^{8} T_{ik} = 1$ for each packet and $\sum_{i=1}^{N_{\mathrm{pkt}}} T_{ik} = \frac{N_{\mathrm{pkt}}}{8}$ for each importance level, ensuring full and balanced assignment. Fig.~\ref{fig:learned_routing_system} illustrates the internal structure of the importance assignment module. During training, hard assignments are used in the forward pass to simulate actual packet transmission, while soft assignments are used in the backward pass to preserve differentiability. Fig.~\ref{fig:packet_channel_assignment} shows an example of the resulting soft and hard assignment matrices.

Since the learned importance assignment module produces assignments that are not deterministically recoverable at the receiver, the packet-to-level assignment matrix must be transmitted as metadata alongside the compressed bitstream. The required overhead is
\begin{equation}
B_{\mathrm{assign}} = N_{\mathrm{pkt}} \log_2(8) \quad \text{bits}.
\end{equation}
For our configuration with $N_{\mathrm{pkt}} = 128$, this corresponds to $384$ bits per frame, which remains negligible compared with the size of the compressed video stream.

\subsection{Multi-Level Erasure Channel Abstraction}
\label{sec:channel_interface}

We abstract the unreliable transmission channel as a multi-level block erasure model used for end-to-end training of the proposed framework. The abstraction consists of $L$ levels, each characterized by a packet erasure probability $\epsilon_l$ for $l \in \{1, \ldots, L\}$, ordered as $\epsilon_1 \leq \epsilon_2 \leq \cdots \leq \epsilon_L,$ such that level 1 offers the greatest erasure protection and level $L$ the least. Each level carries an equal number of packets to preserve balanced utilization.

This abstraction serves as a general model for erroneous channels rather than requiring a physical multi-channel infrastructure between encoder and decoder \cite{tung2024multilevel,nc340417}. By training over this multi-level erasure structure, the framework learns to assign the most important features and spatial regions to the highest-protection levels, ensuring preferential protection of task-relevant content. During inference, the actual channel conditions need not match the training erasure probabilities exactly; as long as the relative ordering of protection levels is preserved, the system maintains robust performance and graceful degradation under varying channel conditions.

Beyond its role in training, this multi-level abstraction aligns naturally with emerging Quality of Service (QoS) standards for wireless transmission. Specifically, the proposed framework is compatible with recent QoS standards in which a protocol data unit (PDU) set contains multiple PDU blocks, equivalent to the packets assigned to a single importance level in our framework, each capable of carrying a different PDU set importance (PSI) level. This compatibility enables direct integration of the proposed semantic-aware codec into existing and next-generation transport and communication architectures without requiring modifications to the underlying protocol stack.

\subsection{Error-Resilient Entropy Coding and Decoding}
\label{sec:decoding}

After importance assignment, each packet is independently entropy coded and transmitted through the erasure channel. At the receiver, successfully received packets are entropy decoded and used to reconstruct the latent representation before being passed to the decoder, and missed or dropped packets are replaced with the masking value zero.

The entropy model plays a critical role in neural video compression. Given an encoded latent representation, the entropy model estimates the probability distribution of each latent symbol and uses this distribution for arithmetic coding. Since arithmetic coding requires identical probability estimates at both the encoder and decoder, even small estimation errors can lead to incorrect bitstream reconstruction and decoding failures. To maintain consistency between the encoder and decoder, a hyperprior autoencoder generates compact side information $\mathbf{z}$ describing the latent distribution. The hyperprior is transmitted together with the encoded bitstream, and at the decoder, the hyper-decoder expands $\mathbf{z}$ back to the original latent size to recover the probability model.

In the original \ac{DCVC}-\ac{RT} architecture, entropy estimation is performed sequentially across packets. Let the latent representation be partitioned into packets $\mathbf{Y} = \{\mathbf{Y}_1, \mathbf{Y}_2, \ldots, \mathbf{Y}_N\}$. The entropy model estimates the distribution of packet $\mathbf{Y}_i$ using not only the hyperprior information but also the previously decoded packets $\mathbf{Y}_1, \mathbf{Y}_2, \ldots, \mathbf{Y}_{i-1}$, i.e., 
$$p(\mathbf{Y}_i) = p\left(\mathbf{Y}_i \mid \mu_1, \sigma_1, \cdots \mu_{i-1}, \sigma_{i-1}, \mu_i, \sigma_i\right).$$
While this sequential strategy improves compression efficiency under reliable communication, it introduces strong inter-packet dependencies. If an earlier packet is lost, the decoder can no longer accurately estimate the probability distribution of subsequent packets, causing error propagation that severely degrades reconstruction quality.

To address this limitation, we redesign the entropy model such that each packet is entropy coded and decoded independently. The single hyperprior $\mathbf{z}$ is still encoded and transmitted as before. At both the encoder and decoder, the hyper-decoder expands $\mathbf{z}$ to the full latent size, producing a global mean and variance map $(\boldsymbol{\mu}, \boldsymbol{\sigma})$ over all spatial locations and feature slices. To obtain the entropy parameters for a specific packet $\mathbf{Y}_i$, we apply a binary mask that selects only the entries of $(\boldsymbol{\mu}, \boldsymbol{\sigma})$ corresponding to the spatial patch and feature slice associated with packet $i$, yielding per-packet parameters $(\mu_i, \sigma_i)$. The probability distribution of each packet is therefore estimated as
$$p(\mathbf{Y}_i) = p\left(\mathbf{Y}_i \mid \mu_i, \sigma_i\right),$$
without dependence on any other packet. The proposed entropy model retains the major components of the original \ac{DCVC}-\ac{RT} framework, including the hyper-encoder, hyper-decoder, and prior-fusion network, with all cross-packet dependencies removed.

\begin{figure}[h]
    \centering
    \includegraphics[width=1\linewidth]{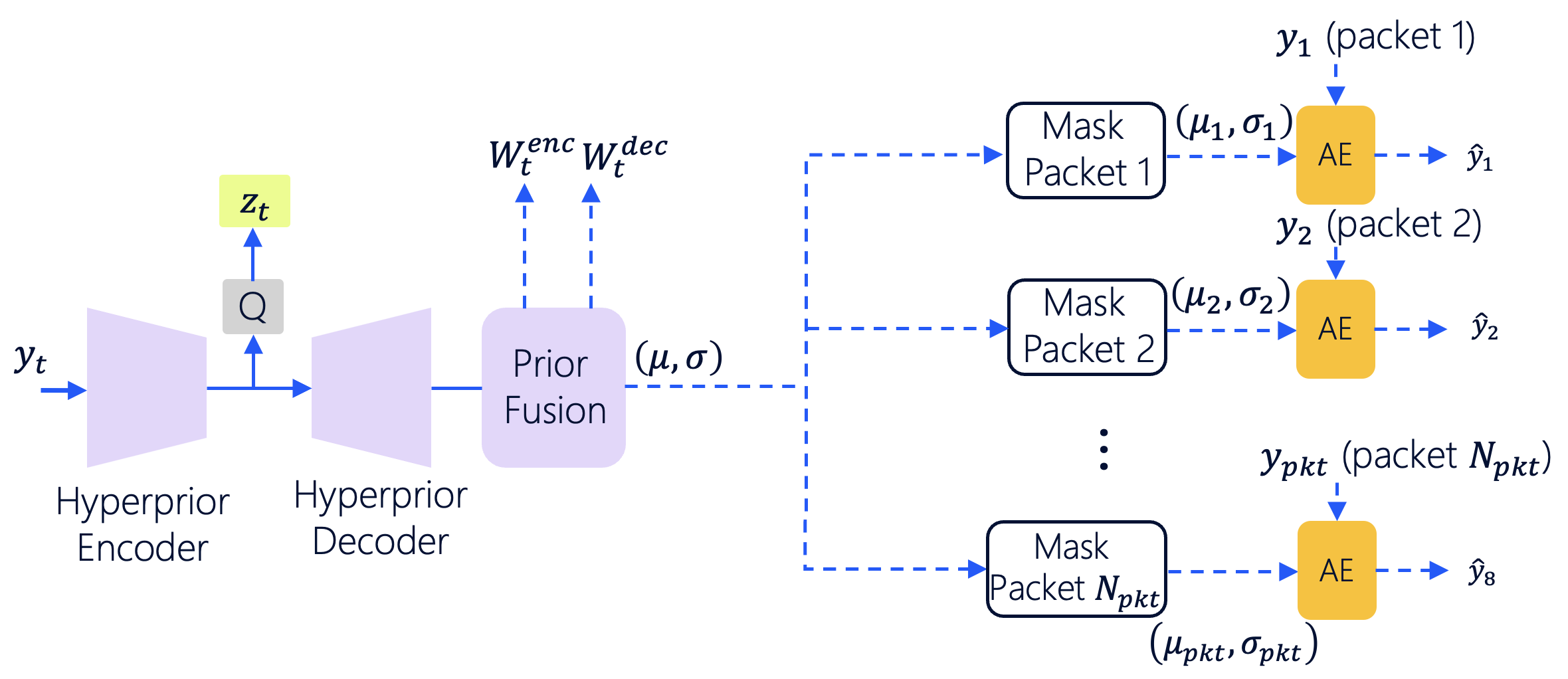}
    \caption{Error-resilient entropy coding and decoding. A shared hyperprior $\mathbf{z}_t$ is expanded and fused to produce a global $(\boldsymbol{\mu}, \boldsymbol{\sigma})$ map, from which per-packet parameters $(\mu_i, 
\sigma_i)$ are extracted via spatial masking. Each packet $y_i$ is then arithmetically encoded and decoded independently, eliminating inter-packet dependencies and enabling graceful degradation under packet erasures.}
    \label{fig:entropy_coding}
\end{figure}

At the receiver, successfully received packets are decoded independently and inserted into their corresponding positions within the latent representation. Missing packets are replaced with zeros before reconstruction. The resulting latent tensor is then processed by the DCVC-RT decoder to generate the reconstructed frame. Since each packet's decoding depends only on the shared hyperprior $\mathbf{z}$ and the packet itself, the loss of any packet does not affect the decoding of others, enabling graceful degradation under erasures.

\subsection{End-to-End Semantic and Erasure-Aware Learning}
\label{sec:training}

The proposed framework is trained end-to-end by jointly optimizing the video encoder, entropy model, importance assignment module, and decoder under stochastic packet erasures. Unlike the original \ac{DCVC}-\ac{RT} framework, which is optimized solely for video compression using a conventional rate-distortion objective
\begin{equation}
\mathcal{L}_{\mathrm{RD}} = D + \lambda R,
\end{equation}
where $D$ denotes the reconstruction distortion, $R$ denotes the bitrate, and $\lambda$ controls the rate-distortion tradeoff, the proposed system explicitly incorporates semantic importance and erasure robustness into the training objective, enabling task-oriented JSCC behavior.

To prioritize task-relevant content, we replace the conventional distortion term with a semantic-weighted MSE loss. Using the pixel-level importance map $\mathbf{M}(\mathbf{x})$ defined in Section~\ref{sec:semantic_map}, the semantic distortion loss is defined as
\begin{equation}
\mathcal{L}_{\mathrm{sem}} = \frac{1}{N} \sum_{\mathbf{x}} \mathbf{M}(\mathbf{x}) \bigl( x - \hat{x} \bigr)^2,
\end{equation}
where $N$ denotes the number of pixels in the frame and $\hat{\mathbf{x}}$ denotes the reconstructed frame. Reconstruction errors in semantically important regions thus incur larger penalties, encouraging the encoder and decoder to preserve task-relevant content with higher fidelity.

Three additional loss terms are introduced to guide the importance assignment module. The first is an assignment entropy loss $\mathcal{L}_{\mathrm{ent}}$ that encourages importance assignments to converge toward deterministic packet-to-level mappings. Let $\mathbf{T}$ denote the soft transport matrix produced by the Sinkhorn normalization layer. This loss is defined as:
\begin{equation}
\mathcal{L}_{\mathrm{ent}} = -\frac{1}{N_{\mathrm{pkt}}} \sum_{i=1}^{N_{\mathrm{pkt}}} \sum_{k=1}^{8} T_{ik} \log T_{ik},
\end{equation}
which penalizes uncertain assignments and promotes near one-hot assignment decisions. The second and third terms are a patch-priority loss $\mathcal{L}_{\mathrm{patch}}$ and a feature-priority loss $\mathcal{L}_{\mathrm{feature}}$, which encourage packets from semantically important regions and important latent feature slices to be assigned to higher protection levels, respectively. Together, these three losses guide the importance assignment module to exploit the multi-level erasure abstraction according to both semantic and feature importance.

Training is performed under stochastic packet erasures, where during each iteration packet losses are randomly generated according to the multi-level erasure abstraction defined in Section~\ref{sec:channel_interface}. The assigned packets are subjected to simulated erasures before entropy decoding and frame reconstruction, exposing the encoder, importance assignment module, and decoder to a wide range of erasure conditions during training.

The overall training objective is
\begin{equation}
\begin{split}
\mathcal{L} &= \mathcal{L}_{\mathrm{sem}} + \lambda_R \mathcal{R} + \lambda_E \mathcal{L}_{\mathrm{ent}} \\
&\quad + \lambda_P \mathcal{L}_{\mathrm{patch}} + \lambda_F \mathcal{L}_{\mathrm{feature}},
\end{split}
\end{equation}
where $\lambda_R$, $\lambda_E$, $\lambda_P$, and $\lambda_F$ are weighting coefficients controlling the relative contributions of bitrate optimization, assignment entropy regularization, patch-priority alignment, and feature-priority alignment, respectively.

\section{Experimental Results}

The proposed framework is trained on the Vimeo-90K dataset~\cite{xue2019video}, which contains 89,800 video clips each consisting of 7 frames. The first frame of each clip is used to train the I-frame model, while the remaining 6 frames are used to train the P-frame model. For P-frame training, the decoded feature representation of the previous frame is extracted and concatenated with the current frame at both the encoder and decoder networks, following the implicit temporal modeling design of \ac{DCVC}-\ac{RT}~\cite{dcvcrt}. During training, a \ac{QP} value is sampled uniformly from the range $[0, 63]$ for each video clip, with all frames within a clip sharing the same QP value. The encoder, decoder, and entropy model architectures follow DCVC-RT~\cite{dcvcrt} exactly, with the proposed framework introducing modifications after the encoding stage: the encoded latent representation is partitioned into packets according to the hybrid feature-region packetization scheme (Section~\ref{sec:packetization}), and the learned importance assignment module allocates each packet to one of the available importance levels based on its semantic and feature importance (Section~\ref{sec:routing}). The semantic importance map generation module is excluded from end-to-end training; randomly generated and spatially smoothed importance maps are used during training as described in Section~\ref{sec:semantic_map}. The entire remaining network, including the encoder, decoder, entropy model, and learned importance assignment module, is trained end-to-end under stochastic packet erasures using the loss function defined in Section~\ref{sec:training}.

For evaluation, we test the system on 5 video sequences of varying resolution ($1280\times720$ and $1920\times1080$) comprising a total of 1262 frames. All reported results are obtained at the highest QP value of 63.

\subsection{Experiment 1: Robustness and Graceful Degradation}

We first demonstrate the ability of the proposed framework to achieve graceful degradation under increasing channel erasures. The model is trained under a uniform erasure probability of $\epsilon = 0.15$ across all 8 channels. During inference, the system is evaluated under a range of uniform erasure probabilities using five long video sequences comprising 1262 frames.

\begin{figure}[t]
    \centering
    \includegraphics[width=0.9\columnwidth]{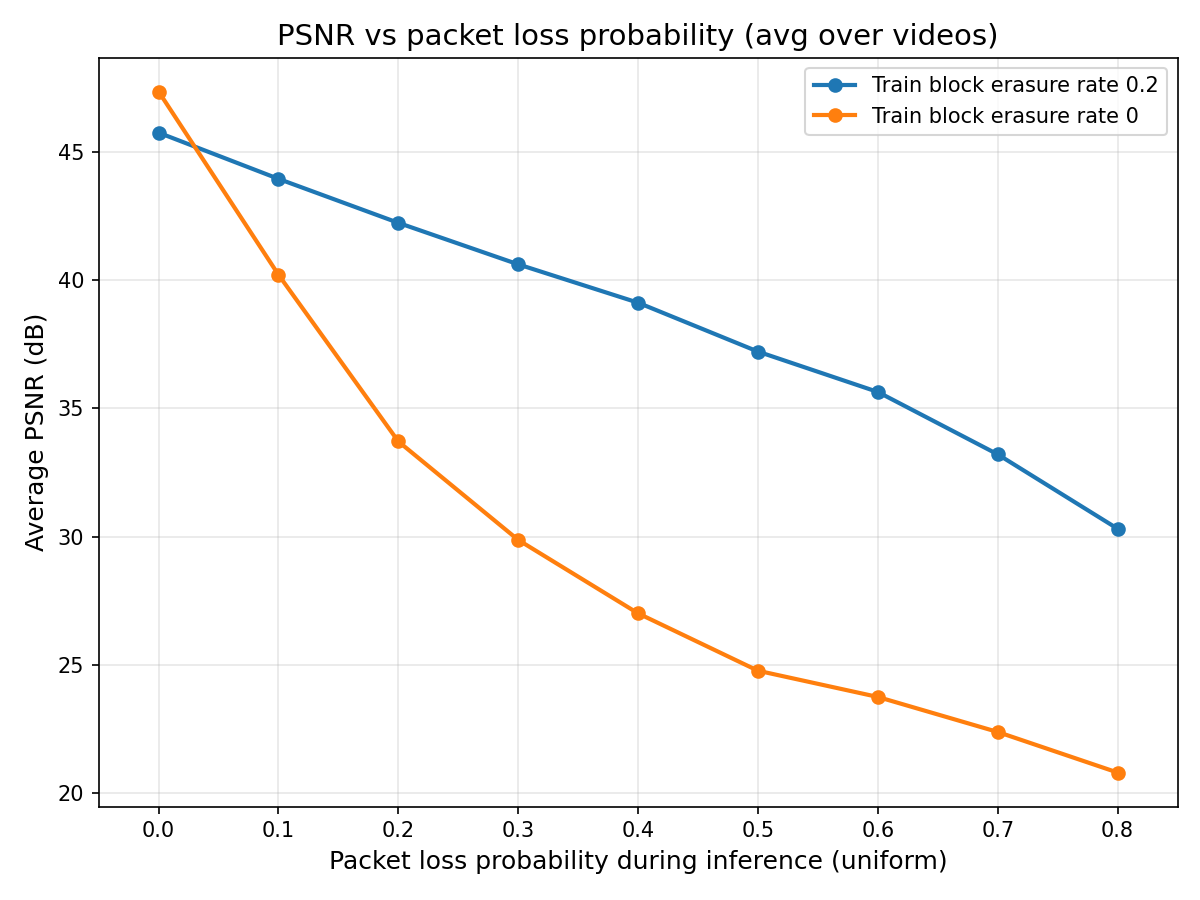}
    \caption{Average PSNR performance across various uniform packet erasure probabilities during inference. The proposed framework demonstrates superior robustness against packet erasures compared to the original DCVC-RT.}
    \label{fig:exp1}
\end{figure}

Fig.~\ref{fig:exp1} shows that the proposed framework consistently achieves higher PSNR compared to the original \ac{DCVC}-\ac{RT} across a wide range of erasure probabilities. Notably, our method maintains superior performance even under channel conditions more severe than those encountered during training, demonstrating robustness and generalization capability. The original\ac{DCVC}-\ac{RT}, due to its sequential entropy dependency, suffers catastrophic degradation even at low erasure rates, whereas our independent entropy model enables graceful quality reduction.

\subsection{Experiment 2: Semantic-Aware Unequal Error Protection}

We evaluate the framework's ability to provide semantic-aware unequal error protection. A semantic segmentation module generates importance maps at the transmitter, which are used by the packet routing module. Training uses non-uniform channel erasure probabilities $\bm{\epsilon}_\text{train} = [0.10, 0.15, 0.18, 0.19, 0.21, 0.22, 0.25, 0.30]$. During inference, more severe conditions are applied: $\bm{\epsilon}_\text{test} = [0.1, 0.2, 0.3, 0.4, 0.5, 0.6, 0.7, 0.8]$.

\begin{figure}[t]
    \centering
    \includegraphics[width=\columnwidth]{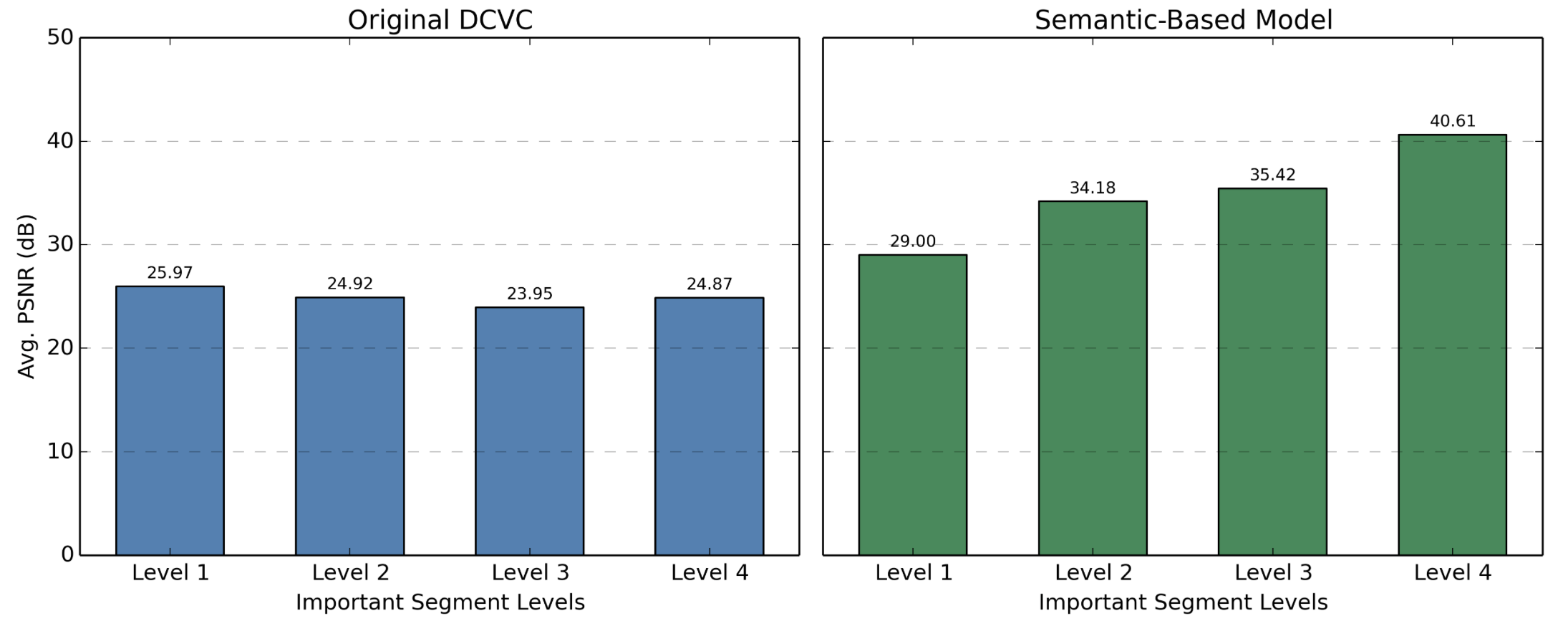}
    \caption{Average PSNR across regions with different semantic importance levels. The proposed framework achieves significantly higher reconstruction quality for semantically important regions compared to the original DCVC-RT which degrades uniformly.}
    \label{fig:exp2}
\end{figure}

Fig.~\ref{fig:exp2} shows that the proposed framework achieves significantly higher reconstruction quality for semantically important regions while gracefully degrading less important regions. In contrast, \ac{DCVC}-\ac{RT} exhibits nearly uniform degradation across all regions. These results confirm that the framework effectively aligns transmission reliability with semantic importance, enabling task-aware communication.

\section{Conclusion}
We presented a semantic-aware, multi-level neural video codec framework for robust, low-latency transmission over unreliable channels. By redesigning the entropy model for independent per-packet decoding, introducing hybrid semantic- and feature-aware packetization, and incorporating a learned importance assignment module that aligns packet importance with protection levels, the proposed system achieves significant improvements in robustness over the baseline under packet erasures. End-to-end training over a multi-level erasure channel abstraction enables JSCC behavior, where the encoder learns representations inherently robust to erasures while the importance assignment module ensures preferential protection of task-relevant content. Experimental results demonstrate graceful quality degradation and effective semantic-aware unequal error protection, making the framework suitable for task-oriented, low-latency video communication over bandwidth-constrained and unreliable wireless channels. 

\bibliographystyle{IEEEtran}
\bibliography{refs}

\end{document}

%% file: abbreviation.tex
\acrodef{URLLC}[URLLC]{ultra reliable low latency communication}
\acrodef{BLER}[BLER]{block error rate}
\acrodef{SNR}[SNR]{signal-to-noise ratio}
\acrodef{LDPC}[LDPC]{low-density parity-check}
\acrodef{AWGN}[AWGN]{additive white Gaussian noise}
\acrodef{DVC}[DVC]{deep video compression}
\acrodef{NVC}[NVC]{neural video codec}
\acrodef{DCVC}[DCVC]{deep contextual video compression}
\acrodef{RT}[RT]{real time}
\acrodef{PSNR}[PSNR]{peak signal-to-noise ratio}
\acrodef{MS-SSIM}[MS-SSIM]{multi-scale structural similarity index}
\acrodef{DeepJSCC}[DeepJSCC]{deep joint source-channel coding}
\acrodef{QP}[QP]{quantization parameter}